\documentstyle[amssymb,aps,twocolumn]{revtex}

\begin{document}
\author{P. Mendels$^{1}$, J. Bobroff$^{1}$, G. Collin$^{2}$, H. Alloul$^{1}$, M.
Gabay$^{1}$, J.F. Marucco$^{3}$, N. Blanchard$^{1}$ and B. Grenier$^{1}$}
\title{Normal State Magnetic Properties of Ni and Zn Substituted in YBa$_{2}$Cu$%
_{3} $O$_{6+x}$: Hole-Doping Dependence}
\address{$^{1}$Lab. de Physique des Solides, UMR 8502, Universit\'{e} Paris-Sud,
91405 Orsay Cedex, France\\
$^{2}$Lab. L\'{e}on Brillouin, CE Saclay, CEA-CNRS, 91191 Gif sur Yvette,
France\\
$^{3}$Lab.des Compos\'{e}s Non Stoechiom\'{e}triques, Universit\'{e}
Paris-Sud, 91405 Orsay Cedex, France}
\date{17 March 1999}

\twocolumn[\hsize\textwidth\columnwidth\hsize\csname@twocolumnfalse\endcsname
\maketitle

\begin{abstract}
We present SQUID susceptibility data on Zn and Ni substituted YBa$_{2}$Cu$%
_{3}$O$_{6+x}$. Cross-checks with NMR yield an unprecedented accuracy in the
estimate of the magnetic susceptibility associated with the substituants,
from the underdoped to the lightly overdoped case. This allows us to
determine the Weiss temperature $\theta $ for YBCO: its value is very small
for all hole dopings $n_{h}$. Since in conventional metals, the Kondo
temperature, $T_{K}<\theta $, magnetic screening effects would not be
expected for $T\gg \theta $; in contrast, increasing $n_{h}$ produces a
reduction of the small moment induced by Zn$^{2+}$ and a nearly constant
effective moment for Ni$^{2+}$ corresponding to a spin 1/2 rather than to a
spin 1.
\end{abstract}

]

There is widespread interest in the study of substitutions in correlated
magnetic systems. It encompasses high $T_{c}$ cuprates, with the well known
depression of $T_{c}$, as well as cuprate-related materials such as spin
ladders and even Cu chain materials. All of these systems share the features
that the local magnetic environment around the defect differs from that of
the bulk material and creates a long range magnetic perturbation. For
instance, a staggered static magnetization is induced by defects in chain
compounds\cite{Takigawa}; similarly in cuprates local moments appear around
Zn substituted in the CuO$_{2}$ planes, as observed by $^{89}$Y NMR\cite
{Mahajan}. Yet, the underlying theory is especially appealing for high $T_{c}
$ cuprates since holes play a dominant role and require a different
microscopic approach. In turn, in this context, the study of substitutions
in the CuO$_{2}$ planes helps to achieve a more complete understanding of
the normal state.

The differences between various substituants for Cu in cuprates, e.g. Zn$%
^{2+}$ ($3d^{10}$, $S=0$) and Ni$^{2+}$ ($3d^{8}$, $S=1$), are still a
matter of debate and probably lie not only in the magnetic properties of
these 2 ions but also in the electronic structure of the defect centres,
which are responsible for the scattering of holes\cite{Ong}. Theories
clearly suggest that, around the impurities, the underlying
antiferromagnetic correlations of the host are revealed\cite{Poilblanc}.
Hole-doping dependence is thus expected and is already observed in $T_{c}$
depression\cite{MendelsGr} and resistivity \cite{Fukuzumi}. On the magnetic
side, the focus of this Letter, accurate determinations of the
susceptibility associated with such defects have been, up to now, prevented
by the existence of spurious magnetic phases\cite{Zagoulaev,Walstedt}.
Thanks to special care in sample preparation, we present here, for the first
time, a full study of the variation with hole-doping of the effects of Ni
and Zn substitutions on the magnetism of the correlated CuO$_{2}$ planes.
Our novel experimental determination of the Weiss-temperature enables us to
demonstrate that, for Zn, the weakness of the moment is not related to the
classical Kondo effect. We present a possible explanation based on strong
correlations.

Of all the cuprates, YBa$_{2}$Cu$_{3}$O$_{6+x}$ is the best suited for such
a study, as the effect of doping on local magnetism can be followed by
merely changing the oxygen content and does not require new synthesis. Our
YBa$_{2}$(Cu$_{1-y}$M$_{y}$)$_{3}$O$_{6+x}$ (M = Zn, Ni) samples, hereafter
noted O$_{6+x}$:M$y\%$, were prepared at 920${{}^{\circ }}$C, using Zn and
Ni oxides to introduce impurities, with $y<$ $4\%$. Above this limit,
substantial amounts ($\sim $1\%) of spurious phases were created.
Deoxidizations have been performed under thermogravimetric control between $%
350{{}^{\circ }}$C and $430{{}^{\circ }}$C in a 0.1 mbar vacuum or with a 10
mbar partial pressure of oxygen for $0.85<x<1.0$ (accuracy $\Delta x=0.01$).
Magnetization data were taken using a SQUID\ magnetometer in 0.2-5 Tesla
fields and $T_{c}$'s were determined from ac susceptibility or SQUID
measurements in low field.

The susceptibility $\chi $ can be divided into 3 contributions, $\chi =\chi
_{m}+\chi _{_{\text{YBCO}}}+\chi _{spur}$. The first one, of primary
interest here, $\chi _{m}$, is associated with local moments induced by
substitution. It will be shown to follow accurately a Curie (or Curie-Weiss)
law. The second one is similar to that of non-substituted YBCO and can be
further split into 2 parts, $\chi _{_{\text{YBCO}}}=\chi _{Cu(2)}+\chi _{ch}$%
, associated respectively with planar Cu far from the substitution site, and
with CuO chains. NMR, as will be shown hereafter, allows one to contrast
between $\chi _{Cu(2)}$ and $\chi _{ch}$. Finally, Curie-like spurious terms
due to extrinsic phases, $\chi _{spur}$ are usually not easily separable
from $\chi _{m}$.

As a prerequisite to the determination of $\chi _{m}$, studies of $\chi _{_{%
\text{YBCO}}}$ were first performed on pure powdered or sintered samples
prepared in the same manner. We succeeded in preparing nearly
parasitic-phase-free samples ($\chi _{spur}\approx 0$) as evidenced by the
absence of any detectable Curie behavior for $x=1.0$ (fig.1). Our accuracy
allows a refined analysis of $\chi _{_{\text{YBCO}_{7}}}$ beyond the
(unjustified) usual assumption that $\chi $ is constant. The intrinsic spin
susceptibility of the planes, measured by $^{89}$Y NMR on our sample, {\it %
decreases} by about 7\% ($\sim 2\times 10^{-8}$ emu/g), from 150 to 300 K,
as commonly observed by NMR \cite{Alloul-Klein},\cite{shiftO-Y-Cu}. Thus the
observed $\sim $1\% quasi-linear {\it increase} of $\chi _{_{\text{YBCO}%
_{7}}}$ over this $T$ range, fitted as $\delta \chi _{_{\text{YBCO}%
_{7}}}=AT, $ results from a near cancellation of the decreasing planar
susceptibility with an increasing chain spin susceptibility ($\chi _{ch}$),
consistent with Cu(1) NMR\ shift measurements\cite{Walstedt2}. The
susceptibility of our set of pure samples, all prepared from the same master
batch, is also reported in the inset of fig.1, for $0.55<x<1.0$. For the
most depleted samples ($x\leqslant 0.65$), $\chi _{Cu(2)}$ is known from $%
^{17}$O NMR to decrease monotonously at low $T$\cite{Yoshinari,Bobroff2}.\
Therefore, the flattening displayed in the inset of fig.1 for $x=0.66,0.55$
and $T$ $<100$ K is not associated with $\chi _{spur}$, as the low-$T$ heat
treatment performed in order to deoxidize our sample cannot create any new
parasitic phases. It is the signature of a minor paramagnetic contribution
due to $\chi _{ch}$, very likely short chain segments. Finally, for $x=1.0$, 
$^{89}$Y NMR\ shift measurements at 100 K on our samples show that Zn
substitution does not affect the level of hole doping in the planes, to
within 0.02 in $x$\cite{Alloul-Mendels}. We have therefore chosen to refer
the $x$ value to $x=1.0$ taken for the maximum obtained oxygen content,
whatever the substitution fraction, $y$.

The smallness of the value of the induced moment in the Zn case makes the
analysis of the data far more difficult than for Ni. We first study a $y=4\%$
Zn substitution for $x=1.0$ or $0.66$. Fig.1 shows our data for the O$_{7.0}$
:Zn1\% and 4\% lightly overdoped samples in comparison with the
corresponding pure sample. The contribution of $\chi _{m}$ can be extracted
by a fit of the $T$- dependent part of the raw data to a $AT+C/T$ law where
the $AT$ term accounts for the contribution of the nonmagnetic Cu sites
(chain + planes) similar to $\delta \chi _{_{\text{YBCO}_{7}}}$, as detailed
in the previous paragraph\cite{note Monod}. There was no significant
improvement in the quality of the fit if we took a Curie-Weiss $C/(T+\theta
) $ behavior ($-5<\theta <10$ K). Fig.2 shows our results for O$_{6.66}$%
:Zn1\% and 4\% underdoped samples. We focus first on the latter for which
data could be extended down to 5K, since it is not superconducting.\ This
makes the Curie increase of $\chi $ at low $T$ more evident. However, the
contribution of $\chi _{_{\text{YBCO}}}(T)$ to the total $T$-variation of $%
\chi _{m}$, is already quite sizeable above 60 K (much more than for $x=1.0$%
). Various estimates of $\chi _{_{\text{YBCO}}}$ were therefore performed,

(i) using the susceptibility of the pure sample, $\chi _{_{\text{YBCO}%
_{6.66}}}$, presented in the inset of fig.1 (dotted line), which limits the
fit to the $T$-range $T>$75 K

(ii) using the data from planar $^{17}$O or $^{89}$Y NMR measurements
rescaled by the hyperfine constant and extrapolated linearly to zero spin
shift at $T=0$.\newline
The main advantage of (i) is to include contributions from the chains, ($%
\chi _{ch}$), at low $T$\cite{note chain} (see fig.1) whereas in (ii) $\chi
_{ch}$ is completely neglected. The local moment contribution, $\chi
_{m}=\chi -(1-5c)\chi _{_{\text{YBCO}}}$ is then deduced, yielding the
correct Curie-Weiss fit over the whole $T$-range. The factor $1-5c$ ($c=1.5y$
is the concentration per plane) reflects the fact that the substituted Cu
site and its four neighbors which carry the local moment are not expected to
have the same susceptibility $\chi _{_{\text{YBCO}_{6.66}}}$, as the bulk Cu
of the CuO$_{2}$ planes (as shown by Y NMR\cite{Mahajan}). The high-$T$ fit
(procedure (i)) yields $C=$ $2.3(2)\times 10^{-5}$ emu.K/g and $\theta =5(3)$
K whereas, including the low-$T$ data gives (ii) $C=$ $2.5(2)\times 10^{-5}$
emu.K/g and $\theta =4(1)$ K. $C$ is modified by less than 15\% if we
extract $\chi _{m}$ by removing the contribution of the substituted site
only, that is ($1-c)\chi _{_{\text{YBCO}}}$. The dominance of the Zn-induced
contribution at low T ($T<30$ K) makes the value of $\theta $ hardly
dependent on the $T=0$ extrapolation performed for $\chi _{_{\text{YBCO}}}$
in approach (ii). It is remarkable that all methods yield very similar
results within a 15\% error bar for a given sample. The value of $\theta $
was further checked by studying the variation of the magnetization $M$ with
the applied field $H$ for $5<T<12$ K (fig.3); indeed, one expects $M$ to
saturate for high fields and low $T$, according to the Brillouin function $%
M=B_{J}(\mu _{B}H/k_{B}(T+\theta ))$. We find $\theta =4.5(5)$ K, in good
agreement with the previous values. Finally, it is interesting to note that $%
\theta $ is quite small and comparable to the transition temperature of the
disordered magnetic state ($\simeq $2 K) observed in $\mu $SR for the same O$%
_{6.66}$:Zn4\% sample\cite{Znmuons}.

Determinations of $C$ were also made along the same lines for lower Zn
substitution rates but the occurrence of superconductivity prevented us from
estimating $\theta $ as accurately as for $y$ = 4\%. In any case the data
plotted in fig.4 clearly demonstrates that $\theta $ never exceeds a few K.

Our complete experimental data analysis for Zn and Ni, partially reported on
fig.1, 2 and 4, is summarized in the inset of fig.4. We find that $C$
increases linearly with $y$. The values of the local moment are deduced from
a linear fit of $C(y)$ $\sim y\mu _{eff}^{2}$ and results are given in Table
I, along with the depression of $T_{c}$. One of our important experimental
conclusions is that the value of the Ni moment stays quasi-constant with
hole doping and near a spin $1/2$ value ($\mu _{eff}=1.73\mu _{B}$), far
from the expected $S=1$ for a Ni$^{2+}$ ion ($\mu _{eff}=2.82\mu _{B}$). In
contrast, the moment induced by Zn is quite small. $C$ decreases by a factor
5.8 from the under- to the lightly over-doped side of the YBCO phase
diagram. One might wonder whether the reduction of the moment for Zn is due
to a classical Kondo effect. In ordinary metals, such a reduction only
occurs below $\theta $, which is of the order of the Kondo temperature $%
T_{K} $. On the contrary, here, for $T\gg \theta $, the moment is reduced
whereas $\chi _{m}$ deviates only slightly from a Curie law (moreover $%
\theta $ is rather associated with long-range interactions between moments).

Finally, we monitored the decrease of the Curie constant, $C$, with hole
doping for the 4\% Zn concentration. In the upper panel of fig.5, we
evidence a strong correlation between $C$ and the residual resistivity $\rho
_{0}$ taken from\cite{Fukuzumi}. This experimental finding suggests that
local moments and scattering are closely related for cuprates, which, again,
is at variance with the case of ordinary metals where non-magnetic
impurities induce scattering without creation of any related magnetism.
Accordingly, models should account not only for the scattering cross-section
but also for the value of the moment induced by the substituting element. We
next present a tentative theoretical interpretation, based on strong
correlations, which relates these two quantities.

First, we consider the Zn case where the impurity can be treated as a spin
vacancy. Simulations for a weakly doped t-J model for cuprates yield a local
moment for Zn\cite{Poilblanc}, which agrees qualitatively with our findings
but, for our experiments, the hole content is far larger. The extension to
higher doping can be treated in the framework of gauge theories which
advocate spinon-holon separation. Using this approach, Nagaosa and Lee\cite
{Nagaosa} were able to account for resistivity data on O$_{6+x}$:Zn\cite
{Fukuzumi}, in the two extreme cases $S=0$ (overdoping) and $S=1/2$
(underdoping). Following Gabay\cite{Gabay}, a description covering the whole
doping range can be obtained. Because Zn is in a $3d^{10}$ configuration, a
singlet is always present on the substituted site.\ According to the t-J
picture, this implies that both holons and spinons are scattered off this
site. Their density is thus depleted within a typical radius, $R_{S}$,
around Zn, or in mean field theory within $\lambda _{TF}$, the Thomas-Fermi
length; thus the spin 1/2 carried by those Cu surrounding the Zn impurity
are not fully screened and a local moment $S$ survives in the vicinity of
Zn. For a hole fraction $\delta $, the spinon number per unit cell is $%
1-\delta $, far from the Zn site. Within a disk of radius $a$ (the lattice
constant) centered on Zn, it is reduced to $(1-\delta )\rho _{s}(a)$ with,
for a featureless Fermi surface, 
\begin{equation}
\rho _{s}(a)=1-\left( \frac{a}{\lambda _{TF}}\right) K_{1}\left( \frac{a}{%
\lambda _{TF}}\right)
\end{equation}
($K_{1}$ is a Bessel function), leading to 
\begin{equation}
S=\frac{1}{2}(1-\delta )(1-\rho _{s}(a)).  \label{estimateS}
\end{equation}
For scattering states in 2D characterized by a wavevector $k_{0}$, $R_{S}$
is related to the planar resistivity, $\rho _{0,2D}$ by $R_{S}=e^{2}\delta
/2k_{0}\hbar \times \rho _{0,2D}/c$. In the gauge picture, the main
contribution to $\rho _{0,2D}$ comes from holons, that is, neglecting band
effects, $k_{0}=\sqrt{2\pi \delta /a^{2}}$. Connecting $R_{S}$ to $\lambda
_{TF}$ gives the variation of the effective moment $\mu _{eff}=g\mu _{B}%
\sqrt{S(S+1)}$ with respect to doping. Standard scattering theory, when
applied in 2D, gives that $R_{S}/\lambda _{TF}\sim \delta ^{-3/2}$\cite
{note3D} so that 
\begin{equation}
\frac{a}{\lambda _{TF}}=\alpha (\frac{\rho _{0,2D}}{c})^{-1}\delta ^{-2}
\end{equation}
From the data of ref.\cite{Fukuzumi} for $\rho _{0,2D}/c$ (expressed in $%
k\Omega /\square .\%$Zn) and Eq.\ref{estimateS}, we find that $\alpha $
=0.050(5) fits satisfactorily our data for $\mu _{eff}(x)$, in fig.5. A more
refined treatment [beyond the scope of this Letter] would include taking
into account the actual shape of the Fermi surface.

The case of a magnetic substitution (Ni) is quite different: no magnetic
vacancy is introduced into the CuO$_{2}$ plane. A theoretical approach
paralleling that for Zn impurity scattering also accounts for resistivity
measurements in the case of Ni doping. In this latter case, scattering is
due to the formation of a singlet bound state on the Ni site. In addition
there remains a $3\,d_{3z^{2}-r^{2}}$ $S=1/2$ spin on Ni, hardly coupled to
the planes, in agreement with NMR experiments\cite{Bobroff}. The magnitude
of the moment is then expected to be independent of doping. Interplane
coupling involving the $\,d_{3z^{2}-r^{2}}$ orbital may also help explain
the experimental observation of a small variation of $S$ with hole doping.

To conclude, we have found a strong contrast between the case of Ni and Zn
substitutions in YBCO. Not only the depression of $T_{c}$ and the value of $%
S $ are drastically different but also the interplay of substitution with
charge doping is now obvious in the case of Zn. This sets strong constraints
on theoretical models by requiring them to link local moments and scattering
by impurities, beyond the classical Kondo picture. We have addressed this
point using a model based on correlations but the problem is still open as, 
{\it e.g.}, the variation of the magnetic correlation length with doping
could also play a role. Deviations from a Curie law are found negligible and
more likely due to a very small spin freezing temperature linked with
interaction between moments rather than classical Kondo behavior. The
smallness of the moment observed in O$_{7}$:Zn raises the fundamental
question: does it persist into the heavily overdoped phase of the cuprates?

We thank P. Monod and S. Zagoulaev for discussions and preliminary ESR tests
of our sample quality.

\begin{table}[tbp] \centering%
\begin{tabular}[t]{ccccc}
& O$_{7}$:Zn & O$_{7}$:Ni & O$_{6.66}$:Zn & O$_{6.66}$:Ni \\ 
$-\delta T_{c}$ (K/\%) & 11.0(1.5) & 3.4(7) & 21(2) & 7.5(1) \\ 
$\mu _{eff}$ ($\mu _{B}$) & 0.40 (5) & 1.2(1) & 1.0(1) & 1.6(1)
\end{tabular}
\caption{$T_c$ depression and local moment for Zn and Ni substitutions in lightly overdoped and underdoped YBCO
\label{moment}}%
\end{table}%

\begin{figure}[tbp]
\caption[1]{ Susceptibility of O$_{7}$:Zn1\% and 4\% samples compared to
pure O$_{7}$(open squares). Notice, for the latter, the increase of $\chi $
with $T$; whereas, the susceptibility of the CuO$_{2}$ planes (Y NMR,
shifted up by 10$^{-7}$ emu/g) decreases (dashed line). Inset $\chi _{_{%
\text{YBCO}_{6+x}}}$, for $x=0.56,0.65,0.76,0.85,1.0$.}
\label{fig.1}
\end{figure}

\begin{figure}[tbp]
\caption[2]{Lower panel: High-$T$ plot of $\chi $, $\chi _{m}$ for O$_{6.66}$%
:Zn1\% and 4\%. The closed symbols stand for raw data ($\chi $), the open
ones for $\chi _{m}$, obtained after subtraction of $\chi _{_{\text{YBCO}}}$
using (i) (see text). The lines are Curie-Weiss fits described in the text.
Upper panel: Low-$T$ plot of $\chi _{m}($YBCO$_{6.66}$:Zn4\%) vs. $%
1/(T+\theta )$ ($\theta =4.0$ K). The line corresponds to the same value of $%
C$ as determined in the lower panel.}
\label{fig.2}
\end{figure}

\begin{figure}[tbp]
\caption[3]{Plot of the magnetization versus $H/(T+\theta)$ for various
values of $\theta$. The upper and bottom panels show the deviations from the
expected universality for a Brillouin function, for unappropriate values of $%
\theta$ ($\theta=2$ and 7 K). The universality is only obtained for a narrow
range of values of $\theta$, $\theta=4.5(5)$K (mid-panel).}
\label{fig.3}
\end{figure}

\begin{figure}[tbp]
\caption[4]{$\chi _{m}$ versus $1/T$ for O$_{6+x}$:Zn$y\%$,Ni$y\%$. The
lines are fits described in the text. Inset: Curie constant (in emu.K/g)
versus (\%) impurity content for $x=7.0$ and $x=6.66$. }
\label{fig.4}
\end{figure}

\begin{figure}[tbp]
\caption[5]{Top panel: Evolution of $C$ for $y=4\%$ (closed squares), and $%
\rho_{0,2D}$\protect\cite{Fukuzumi} (open squares) with oxygen content.
Bottom panel: Evolution of $\mu _{eff}$ (closed squares) and $T_{c} $
(triangles), for $y=4\%$. The open squares are estimates for $\mu _{eff}$
using eq. \ref{estimateS} for $S$.}
\label{fig.5}
\end{figure}


\begin{references}
\bibitem{Takigawa}  M. Takigawa {\it et al.,} Phys. Rev. B. {\bf 55}, 14129
(1997).

\bibitem{Mahajan}  A.V. Mahajan {\it et al.,} Phys. Rev. Lett. {\bf 72},
3100 (1994).

\bibitem{Znmuons}  P. Mendels {\it et al.,} Phys. Rev. B. {\bf 49}, 10035
(1994).

\bibitem{Ong}  T.R. Chien, Z.Z. Wang and N.P. Ong, Phys. Rev. Lett., {\bf 67}%
, 2088 (1991).

\bibitem{Poilblanc}  D. Poilblanc, D.J. Scalapino and W. Hanke, Phys. Rev.
Lett. {\bf 72}, 884 (1994); Phys.\ Rev. B. {\bf 50}, 13020 (1994).

\bibitem{MendelsGr}  P. Mendels {\it et al.}, Physica C {\bf 235-240}, 1595
(1994).

\bibitem{Fukuzumi}  Y. Fukuzumi {\it et al}, Phys. Rev. Lett., {\bf 76}, 684
(1996).

\bibitem{Zagoulaev}  S. Zagoulaev, P. Monod and J. J\'{e}goudez, Phys. Rev B 
{\bf 52, }10474 (1995).

\bibitem{Walstedt}  R.E. Walstedt {\it et al., }Phys. Rev. B {\bf 48}, 10646
(1993).

\bibitem{Nagaosa}  N. Nagaosa and P.A. Lee, Phys. Rev. Lett., {\bf 79}, 3755
(1997).

\bibitem{Alloul-Klein}  H. Alloul {\it et al.}, Phys. Rev. Lett., {\bf 70},
1171 (1993). M. Takigawa, W.L. Hults and J.L. Smith, Phys. Rev. Lett., {\bf %
71}, 2650 (1993).

\bibitem{shiftO-Y-Cu}  The small decrease of the Cu(2) spin susceptibility
with increasing $T$ for optimally and lightly over-doped cuprates is
universal. It has been observed through $^{17}$O and $^{89}$Y NMR, see e.g. 
\cite{Alloul-Klein,Walstedt2,Bobroff2}.

\bibitem{Walstedt2}  R.E. Walstedt {\it et al.}, Phys.\ Rev. B,{\bf \ 45},
8074 (1992), fig.6. Using the Mila and Rice analysis, Physica C, {\bf 157},
561 (1989), one gets $\delta \chi _{ch}=$ +1-3.7 $\times 10^{-8}$ emu/g,
whereas $\delta \chi _{planes}=-2.0(6)\times 10^{-8}$ emu/g.

\bibitem{Yoshinari}  Y. Yoshinari {\it et al.,} J. Phys. Soc. Jpn., {\bf 59}%
, 3698 (1990).

\bibitem{Bobroff2}  J.Bobroff {\it et al.}, Phys. Rev. Lett. {\bf 78}, 3757
(1997).

\bibitem{Alloul-Mendels}  H. Alloul {\it et al.}, Phys. Rev. Lett. {\bf 67},
3140 (1991).

\bibitem{note Monod}  $A$ varies slightly from batch to batch in pure YBCO$%
_{7}$, so it was left free in the fit. The values found, $A=$ $3.5-5\times
10^{-11}$emu/K, for all our Zn samples are basically within the limits of
data scatter for pure YBCO. The difference between our results and those of%
\cite{Zagoulaev} lies in the presence of this $T-$linear term which was
masked there by a Curie term due to parasitic phases.

\bibitem{note chain}  $\chi _{ch}$ is slightly underestimated because of the
0.7 multiplication factor.

\bibitem{Gabay}  M. Gabay, Physica C {\bf 235-240},1337 (1994). Note that he
equation in p.1338 gives the local spinon density rather than the local
spin, as mistakenly stated there.

\bibitem{note3D}  This result comes from a 2D extension of the standard 3D
calculation, which yields $\lambda _{TF}/\sqrt{R_{S}}\sim $ constant.

\bibitem{Bobroff}  J. Bobroff {\it et al.}, Physica C {\bf 282-287}, 1389
(1997).
\end{references}
\end{document}